\documentclass[superscriptaddress,aps,pra,floatfix,11pt,nofootinbib]{revtex4-2}

\usepackage{amsmath}
\usepackage{amsfonts}
\usepackage{lipsum}
\usepackage{graphicx}
\usepackage{xspace}
\usepackage{epsf}
\usepackage{setspace}
\usepackage[margin=1in]{geometry}
\usepackage[version=4]{mhchem}
\usepackage{array} 
\usepackage{booktabs}
\usepackage{makecell}
\usepackage{float}
\usepackage{xcolor}
\usepackage{wrapfig}
\usepackage{tikz}
\usepackage{tikz-qtree}
\usepackage{braket}
\usepackage{tabularx}
\usepackage{algorithm}
\usepackage{algpseudocode}
\usepackage{booktabs}
\usepackage{microtype}
\usepackage{booktabs} 
\usepackage{here}

\usepackage{setspace}
\usepackage[]{siunitx}
\usepackage[]{microtype}

\usepackage{stmaryrd}
\SetSymbolFont{stmry}{bold}{U}{stmry}{m}{n}

\definecolor{lightblue}{rgb}{0.63, 0.74, 0.78}
\definecolor{seagreen}{rgb}{0.18, 0.42, 0.41}
\definecolor{orange}{rgb}{0.85, 0.55, 0.13}
\definecolor{silver}{rgb}{0.69, 0.67, 0.66}
\definecolor{rust}{rgb}{0.72, 0.26, 0.06}
\definecolor{purp}{RGB}{68, 14, 156}

\colorlet{lightrust}{rust!50!white}
\colorlet{lightseagreen}{seagreen!50!white}
\colorlet{lightorange}{orange!25!white}
\colorlet{lightlightblue}{lightblue}
\colorlet{lightsilver}{silver!30!white}
\colorlet{darkorange}{orange!75!black}
\colorlet{darksilver}{silver!65!black}
\colorlet{darklightblue}{lightblue!65!black}
\colorlet{darkrust}{rust!85!black}
\colorlet{darkseagreen}{seagreen!85!black}

\definecolor{RYB1}{rgb}{0.72, 0.26, 0.06}%
\definecolor{RYB2}{rgb}{0.18, 0.42, 0.41}%
\definecolor{RYB3}{rgb}{0.63, 0.74, 0.78}%
\definecolor{RYB4}{RGB}{251,220,127}
\definecolor{RYB5}{rgb}{0.69, 0.67, 0.66}
\definecolor{RYB6}{rgb}{0.85, 0.55, 0.13}
\definecolor{RYB7}{RGB}{128, 177, 211}

\definecolor{custompurple}{rgb}{0.578125, 0.40234375, 0.73828125}
\definecolor{custompurple1}{rgb}{0.7734375, 0.77734375, 0.87890625}
\definecolor{custompurple2}{rgb}{0.2734375, 0.046875, 0.51171875}

\definecolor{customgray}{rgb}{0.69921875, 0.69921875, 0.69921875}
\definecolor{customgray2}{RGB}{217,217,217}

\definecolor{customred}{rgb}{0.80859375, 0.1484375, 0.15234375}
\definecolor{customred1}{RGB}{197, 58, 50}
\definecolor{customred2}{RGB}{229, 115, 115}
\definecolor{customred3}{RGB}{239, 154, 154}
\definecolor{customred4}{RGB}{255, 205, 210}

\definecolor{customblue}{rgb}{0.12109375, 0.46484375, 0.703125}
\definecolor{customblue1}{RGB}{58,94,158}
\definecolor{customblue2}{RGB}{102,144,190}
\definecolor{customblue3}{RGB}{153,186,213}

\definecolor{customorange1}{RGB}{224,163,77}
\definecolor{customorange2}{RGB}{232,186,121}

\definecolor{customrust1}{RGB}{198,103,63}
\definecolor{customrust2}{RGB}{212,141,111}

\definecolor{customseagreen1}{RGB}{88,137,135}
\definecolor{customseagreen2}{RGB}{130,166,155}

\definecolor{customdarkpurp}{RGB}{69,43,84}
\definecolor{customdarkpurp1}{RGB}{126,81,122}
\definecolor{customdarkpurp2}{RGB}{216,159,170}
\definecolor{customdarkpurp3}{RGB}{237,209,203}

\usepackage{hyperref}
\hypersetup{
  colorlinks=true,
}

\usepackage[labelfont=bf,justification=justified,singlelinecheck=false,font=small]{caption}

\usepackage[nameinlink]{cleveref}
\crefname{equation}{}{}

\usepackage{empheq}

\usepackage[parfill]{parskip}

\usepackage{bm}
\newcommand{\ve}[1]{\bm{#1}}

\newcommand{\bA}{\ve{A}}

\newcommand{\btheta}{\vec{\boldsymbol{\theta}}}

\DeclareMathOperator{\sgn}{sgn}

\usepackage{titlesec}

\titleformat{\section}
  {\normalfont\large\bfseries} %
  {\thesection}{1em}{}          %

\titleformat{\subsection}
  {\normalfont\normalsize\bfseries}
  {\thesubsection}{1em}{}

\renewcommand\thesection{\arabic{section}}
\renewcommand\thesubsection{\arabic{section}.\arabic{subsection}}

\date{}
\makeatletter
\def\@date{}
\makeatother

\begin{document}

\hypersetup{
  linkcolor=darkrust,
  citecolor=seagreen,
  urlcolor=darkrust,
  pdfauthor=author,
}

\title{Hadamard Random Forest for Reconstructing Real Valued Pure Quantum States}

\author{Zhixin Song}
\email{zsong300@gatech.edu}
\affiliation{School of Physics, Georgia Institute of Technology, Atlanta, GA, USA \vspace{-0.125cm}}

\author{Hang Ren}
\affiliation{Berkeley Center for Quantum Information and Computation, Berkeley, CA, USA \vspace{-0.125cm}}

\author{Melody Lee}
\affiliation{School of Computational Science \& Engineering, Georgia Institute of Technology, Atlanta, GA, USA \vspace{-0.125cm}}

\author{Bryan Gard}
\affiliation{CIPHER, Georgia Tech Research Institute, Atlanta, GA, USA \vspace{-0.125cm}}

\author{Nicolas Renaud}
\affiliation{Netherlands eScience Center, Matrix Building III, Amsterdam, The Netherlands \vspace{-0.125cm}}
\affiliation{Quantum Application Lab, Science Park, Startup Village, Amsterdam, The Netherlands \vspace{-0.125cm}}

\author{Spencer H.\ Bryngelson}
\affiliation{School of Computational Science \& Engineering, Georgia Institute of Technology, Atlanta, GA, USA \vspace{-0.125cm}}
\affiliation{Daniel Guggenheim School of Aerospace Engineering, Georgia Institute of Technology, Atlanta, GA, USA \vspace{-0.125cm}}
\affiliation{George W.\ Woodruff School of Mechanical Engineering, Georgia Institute of Technology, Atlanta, GA, USA}

\begin{abstract}
\begin{center}
    \textbf{Abstract}
\end{center}
Many potential quantum computing applications, such as linear system solvers, require efficient readout to preserve the algorithmic speedup, but only involve real amplitudes.
Conventional tomography methods demand exponential resources for general complex-valued states. 
We introduce a readout method for real-valued pure quantum states based on amplitude superposition with Hadamard operations and a random forest algorithm for sign determination.
Our method reduces the measurement settings required for state vector reconstruction to $\mathcal{O}(N_\mathrm{q})$ for an $N_\mathrm{q}$~qubit system; the post-processing cost remains exponential $\Omega(2^{N_\mathrm{q}})$.
We experimentally validate our method on up to 10 qubits using the latest available IBM quantum processor and demonstrate that it accurately extracts key properties, such as magic.
Our method also outperforms the standard SWAP test for state overlap estimation.
This calculation resembles numerical integration in certain cases and can be applied to extract nonlinear properties, which are important in application fields.
We further implement the method to read out the solution from a quantum linear solver.
\end{abstract}

\maketitle

{\renewcommand{\thefootnote}{}%
\footnotetext{Published version: Z.~Song, H.~Ren, M.~Lee, B.~Gard, N.~Renaud, and S.~H.~Bryngelson, ``Hadamard random forest for reconstructing real valued pure quantum states,'' \href{https://doi.org/10.1038/s42005-026-02717-1}{\textit{Commun.\ Phys.}\ \textbf{9}, 333 (2026)}, doi:\href{https://doi.org/10.1038/s42005-026-02717-1}{10.1038/s42005-026-02717-1}.}}

\section{Introduction}\label{s:intro}

Efficient linear system solvers play a vital role in modern scientific computing and engineering challenges, from accelerating the training of large-scale machine learning models~\citep{liu2024towards} to sophisticated fluid dynamics simulations~\citep{li2025potential}.
Quantum approaches to linear system solvers have attracted attention for their potential to outperform classical methods~\citep{hestenes1952methods,saad1986gmres}.
State-of-the-art quantum linear system algorithms (QLSA)~\citep{childs2017quantum, costa2022optimal, low2024quantum, morales2024quantum} promise super-polynomial speedups in the fault-tolerant regime, where resource-intensive quantum error correction protocols and data encoding schemes such as block-encoding are required. 
Despite these advances, the quantum readout problem~\citep{aaronson2015read} may still limit the performance of many claims of potential quantum advantage.
When studying quantum many-body systems, desired state properties can be extracted efficiently, including magnetization, correlation functions, and entanglement entropy~\citep{huang2020predicting}.
However, in many other scientific problems where QLSA is favorable, the desired properties are nonlinear and involve non-Hermitian observables, making it challenging to construct a measurement scheme even using the powerful classical shadow tomography~\citep{huang2022learning}.
The typical approach is to reconstruct the state using tomographic methods or amplitude estimation~\citep{grinko2021iterative,manzano2023real}, where only a few dominant amplitudes are needed.
Then, apply post-processing to extract the relevant properties~\citep{penuel2024feasibility,patterson2025measurement}.

Efficient quantum state reconstruction is a central task in quantum information science, enabling characterization of quantum states and modeling of quantum devices~\citep{blume2025quantum, ren2019modeling, shaffer2023sample}.
The standard full quantum state tomography (FQST) method~\citep{leonhardt1995quantum} requires an exponential number of measurement settings to iterate over the Pauli basis $\{\sigma_x, \sigma_y, \sigma_z\}$ in each qubit and hence reconstruct a complex-valued density matrix.
Moreover, post-processing typically relies on resource-intensive maximum-likelihood estimation to ensure that the reconstructed state $\tilde{\rho}$ is physical~\citep{hradil1997quantum}.

Approaches have been developed to address the challenges in state tomography. 
Some methods reduce measurement costs by exploiting the low-rank (pure-state) structure of density matrices using compressed sensing~\citep{gross2010quantum} or by leveraging limited entanglement structures~\citep{cramer2010efficient}. 
More recent techniques combine parallel measurements to further reduce the required resources, as seen in quantum overlapping tomography~\citep{cotler2020quantum, yang2023experimental} and tensor network learning approaches~\citep{kurmapu2023reconstructing, guo2024quantum, hu2024experimental}.
However, these methods rely on assumptions about the quantum state's structure or involve additional learning processes with unclear post-processing scaling. 

We show that for QLSA solvers involving only real-valued states, reconstructing those states requires a linear number of measurement settings rather than an exponential overhead.
We want to emphasize that our method targets pure-state reconstruction, whereas tomographic methods typically assume a generic mixed state.
To our best knowledge, the compressed sensing tomography introduced in~\cref{s:method} is the most relevant baseline for pure state reconstruction.
Real-valued states arise naturally in many other quantum algorithms, including Grover's search~\citep{grover1996fast} and the Bernstein--Vazirani algorithm~\citep{bernstein1993quantum}.
With carefully injected resource states, one can achieve universal computation with real-valued states~\citep{delfosse2015wigner}.

Similar ideas have been explored for real-valued observables in classical shadow tomography~\citep{west2025real}. 
This work introduces the \textit{Hadamard Random Forest} (HRF), a technique for reconstructing real-valued pure quantum states.
By restricting the relative phase from a complex number to $\{\pm 1\}$, HRF resolves the phases exclusively in the $\sigma_x$ basis combined with a specialized random forest algorithm~\citep{ho1995random}. 
We validate HRF on IBM's quantum hardware using up to 10~qubits and compare its performance with that of tomographic methods.
In these experiments, HRF outperforms tomographic methods in reconstruction accuracy and runtime. 
HRF accurately estimates state properties from near-term hardware sampling results, including quantum magic and state overlap.
We implement the method in a variational quantum linear system solver~\citep{bravo2023variational}.
These results support the practical application of HRF on quantum hardware.

\section{Methods}\label{s:method}

\subsection{Reconstructing quantum states}\label{ss:state-reconstruction}

\begin{figure}[htb]
    \centering
    \includegraphics[]{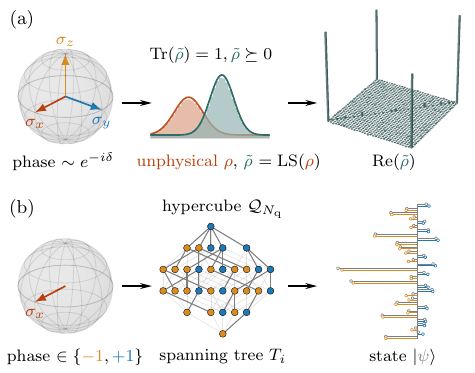}
    \caption{
        Schematic comparison of Full Quantum State Tomography (FQST) and the Hadamard Random Forest (HRF) method for a 5-qubit case.
        (a) FQST requires measuring each qubit in the Pauli basis, resulting in $3^{5}=243$ measurement settings.
        The raw statistics illustrated by the rust curve might be corrupted by experimental noise and lead to unphysical states.
        An optimization procedure is usually required to ensure the reconstructed state is physical with unit trace and positive semi-definite, as in the green curve.
        For pure states with $r=1$, CS--QST empirically requires $80$ settings to match FQST's accuracy.
        (b) HRF only requires $6$ measurement settings and leverage a random forest algorithm to determine the phase on each amplitude (node in the spanning tree).  
    }
    \label{f:overview}
\end{figure}

Given an $N_\mathrm{q}$~qubit density matrix $\rho$, it can be expanded as,
\begin{gather}
    \rho=\frac{1}{2^{N_\mathrm{q}}} \sum_{i, j,\ldots \ell=0}^3 c_{i j \ldots \ell} \underbrace{\sigma_i^{(1)} \otimes \sigma_j^{(2)} \ldots \otimes \sigma_\ell^{(N_\mathrm{q})}}_{N_\mathrm{q} \text{ qubit global observable}}
    = \frac{1}{2^{N_\mathrm{q}}} \sum_{\alpha} c_\alpha P_{\alpha},
    \label{e:quantum-state-tomography}
\end{gather}
where $c_{i j \ldots \ell} \in \mathbb{C}$ and each $\sigma_0=I, \sigma_{1,2,3}= \sigma_{x,y,z}$.
After collecting $3^{N_\mathrm{q}}$ global measurement statistics (each contains $2^{N_\mathrm{q}}$ outcomes), the post-processing involves a convex optimization to ensure the reconstructed state $\tilde{\rho}$ is physical, i.e., positive semi-definite with unit trace as illustrated in \cref{f:overview}.
Optimization is usually formalized as least-squares (LS) fitting. 
For a rank-$r$ state, one can further adopt compressed sensing quantum state tomography (CS--QST)~\citep{gross2010quantum} to reduce the settings to $\mathcal{O}(r \, 2^{N_\mathrm{q}}\, \mathrm{poly}(N_\mathrm{q}))$ random Pauli observables $\{P_\alpha\}$ combined with a trace-norm regularization
\begin{gather}
    \min _{\tilde{\rho}} \sum_{\alpha} \left(\operatorname{Tr}\left(P_\alpha \tilde{\rho}\right)-\operatorname{Tr}\left(P_\alpha \rho\right)\right)^2 + \lambda \operatorname{Tr}(\tilde{\rho}), 
    \quad \text{subject to} \quad 
    \tilde{\rho} \succeq 0,
    \label{e:qst-cs}
\end{gather}
where $\lambda$ is a tunable weight and usually chosen according to the number of measurement settings and sample size~\citep{flammia2012quantum}.
For real-valued pure states ($r=1$), one can reconstruct the state vector $|\tilde{\psi}\rangle$ by taking the $n$-th column of $\tilde{\rho}$ and rescaling it with the square root value of its $n$-th diagonal element.

\begin{figure}[b]
    \centering
    \includegraphics[]{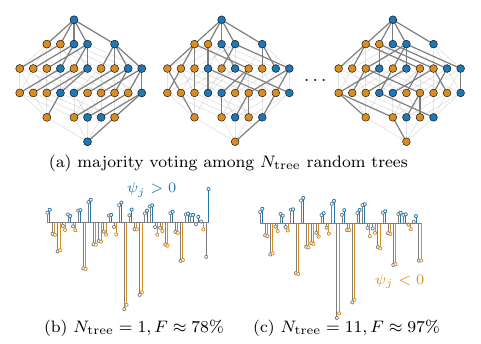}
    \caption{
        Demonstration of using multiple spanning trees to form a random forest and deploy a majority voting scheme to determine the signs.
        (a) Spanning trees in $\mathcal{Q}_5$.
        The dark blue color is $\sgn(\psi_j)=+1$ and light orange is $\sgn(\psi_j)=-1$.
        (b) The last three signs are incorrectly determined for a single tree. 
        (c) No sign error when using 11~trees.
    }
    \label{f:random-forest}
\end{figure}

Consider an $N_\mathrm{q}$~qubit real-valued pure state
\begin{gather}
    \ket{\psi} = U_{\mathrm{prep.}} |0\rangle^{\otimes N_\mathrm{q}} = \sum_{j=0}^{2^{N_\mathrm{q}}-1} \psi_j\ket{j},
    \label{e:real-state-preparation}
\end{gather}
where $U_{\mathrm{prep.}}$ is the unitary corresponding to the state preparation circuit and $\psi_j=\sgn(\psi_j)\cdot|\psi_j| \in \mathbb{R}$.
The basis states $\ket{j}$ are ordered according to the lexicographical order of their bit-string representations.
HRF reconstructs the corresponding state vector by sampling $N_\mathrm{q}+1$ circuits.
The first circuit measures in the standard computational $\sigma_z$ basis, evaluating the amplitudes $|\psi_j|$ by sampling the probability distribution $\{|\psi_j|^2\}$.
Then, one measures each qubit in the $\sigma_x$ basis, resulting in an additional $N_\mathrm{q}$ circuits to sample, which can be achieved by adding a Hadamard gate to rotate the measurement basis for qubit ordering and circuit construction (see Supplementary Note \ref{sn:hrf}).
Applying a Hadamard gate on the $(N_\mathrm{q}-k-1)$-th qubit and sampling the output leads to the new amplitude vector $\ket{\psi^k}$, where $\psi^k_{j}$ and $\psi^k_{j+2^k}$ contain the superposition of original amplitudes as
\begin{equation}
    \psi^k_{j, j+2^k} = \frac{1}{\sqrt{2}}\left(\psi_j \pm \psi_{j+2^k}\right).
    \label{e:measure-statistics}
\end{equation}
This computation leads to $\left(2^{N_\mathrm{q}}\right) \left(N_\mathrm{q}+1\right)$ total probabilities.
One determines the relative signs between two amplitudes $s_{j,j+2^k} \coloneqq \sgn (\psi_j \psi_{j+2^k})$ by querying the sampling results $\{|\psi^k_j|^2\}_{k=0}^{N_\mathrm{q}-1}$ as
\begin{gather}
    s_{j,j+2^k} = \textnormal{sgn} {\big[}\underbrace{2|\psi^k_j|^2}_{\text{new prob.}} - \underbrace{|\psi_j|^2 - |\psi_{j+2^k}|^2}_\text{original prob.} \, {\big]}.
    \label{e:sign-determination}
\end{gather}
All $2^{N_\mathrm{q}}$ signs ($s_j\coloneqq\sgn(\psi_j)$) follow by tracing the path from the root amplitude $\psi_0$ (assuming $\psi_0>0$) to the nodes $\psi_j$ and repeatedly applying the sign determination procedure of \cref{e:sign-determination}.
Identifying such paths is equivalent to finding spanning trees $T_i$ on the hypercube graph $\mathcal{Q}_{N_\mathrm{q}}$ with $2^{N_\mathrm{q}}$ nodes and $N_\mathrm{q}2^{(N_\mathrm{q} - 1)}$ edges.
The edge exists only when the node index differs by a power of $2$.
For example, $\mathcal{Q}_5$ contains $80$ edges (light gray) as illustrated in \cref{f:random-forest}~(a).

The hypercube graph $\mathcal{Q}_{N_\mathrm{q}}$ has at least $\Omega(2^{2^{N_\mathrm{q}}-N_\mathrm{q}-1})$ spanning trees~\citep{harary1988survey}.
The tree structure follows Pascal's triangle using a breadth-first search (BFS) approach for computational efficiency.
Consequently, node $j$ belongs to layer $k$ if its binary representation contains exactly $k$ ones and layer $k$ contains $\binom{N_\mathrm{q}}{k}$ nodes with edges connecting only nodes in adjacent layers. 
First, the tree is initialized with root node $0$ (the unique layer-$0$ node) and connected to all $N_\mathrm{q}$ power-of-two nodes $2^0, 2^1, \ldots, 2^{N_\mathrm{q}-1}$, which constitute layer 1. 
We then perform a breadth-first traversal through layers $k = 2, 3, \ldots, N_\mathrm{q}$, where each node in layer $k$ selects uniformly at random one parent from its $k$ neighbors in layer $k-1$ that are already present in the tree. 
This construction guarantees a valid spanning tree because every layer-$k$ node has at least one neighbor in layer $k-1$ (obtained by flipping any of its $k$ set bits).
The random parent selection ensures that independent executions yield structurally diverse trees, a property essential to the efficacy of majority voting in sign determination. 
This structure also ensures the tree depth is bounded by $L = N_\mathrm{q}$.
We collect the sign determination results of $N_{\mathrm{tree}}$ randomly generated spanning trees, then employ a majority voting scheme with a vote threshold of $50\%$.
The full HRF algorithm is included in Supplementary Note \ref{sn:hrf}.

The pre-processing step is independent of the quantum state structure, including rank and sparsity, and only depends on the system size $N_\mathrm{q}$. 
The tree generation time has an exponential cost as the minimal operation on a graph is $\Omega{(V)}$, where $V$ is the number of nodes. 
While generating the tree has an exponential cost, it is a one-time computation. 
Once constructed, the tree can be cached and reused for any $N_\mathrm{q}$~qubit configuration.
Moreover, since the hypercube graph is always bipartite, we can recursively grow a larger tree out of many smaller trees constructed in parallel.
We benchmark this strategy for larger problem sizes on an Apple M4 Pro chip with 24 \unit{\giga\byte} unified memory.
The recursive tree generation can achieve $2\times$ savings on computation time and $5\times$ for memory usage as shown in \cref{tab:tree-bench}.
\begin{table}[htb]
    \centering
    \begin{tabular}{@{} c|  r  r | r  r @{}}
        \toprule
        Problem size $N_\mathrm{q}$ & BFS time (\unit{\second}) & Recursive time (\unit{\second}) & BFS memory (\unit{\mega\byte})  & Recursive memory (\unit{\mega\byte}) \\
        \midrule
        10  &  0.35  &  0.16  &  5.20  &  1.21  \\
        12  &  1.39  &  0.75  &  20.53  &  5.01  \\
        14  &  6.43  &  3.41  &  108.49  &  19.89  \\
        16  &  29.53  &  15.20  & 456.35  &  80.13  \\
        18  &  146.72  &  72.24  & $>$1000.0  &  321.03   \\
        \bottomrule
        \end{tabular}
    \caption{
        Benchmark results on BFS tree generation time and memory usage for $N_{\mathrm{tree}}=111$. \\
        The recursive method starts with $N_\mathrm{q}=8$.
    }
    \label{tab:tree-bench}
\end{table}
\subsection{Sampling cost and error analysis}\label{ss:sample-cost}

\begin{figure}[htb]
    \centering
    \includegraphics[]{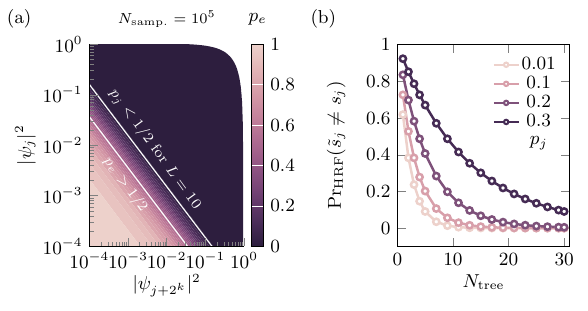}
    \caption{
    Analysis of sign errors at node and tree level.
    (a) Probability $p_e$ of returning a sign error between node $j$ and $j+2^k$ using $10^5$ samples, where the variables are constrained with $|\psi_{j}|^2+ |\psi_{j+2^k}|^2 \leq1$. 
    For 10-qubit states ($L=10$), it requires $p_e< 0.05$ such that $p_j <1/2$ and each tree behaves better than a random guess.
    (b) Probability of estimating the sign $s_j$ wrong after majority voting with different $p_j$ and $N_{\mathrm{tree}}$.
    }
    \label{f:error-bound}
\end{figure}

Next, we analyze the error bounds of using different sample sizes $N_\mathrm{samp.}$ and random trees $N_{\mathrm{tree}}$. 
First, we estimate the probability of a single sign error between $\psi_j$ and $\psi_{j+2^k}$ is $\mathrm{Pr}\left(\tilde{s}_{j,j+2^k}\neq s_{j,j+2^k}\right) \leq \exp{(-2N_\mathrm{samp.}|\psi_j|^2|\psi_{j+2^k}|^2)}$ according to the Hoeffding's inequality~\citep{hoeffding1963probability}, where $\tilde{s}$ is the estimated sign using HRF.
\Cref{f:error-bound}~(a) shows that the larger the product $\psi_{j}\psi_{j+2^k}$, the easier it becomes to resolve their relative sign with fewer shots.
Then, one can analyze the error propagation on each tree.
Since every sign at node $j$ is determined by a unique path of edges from the root $\psi_0$ of length $d_j\le L$, the estimated $\tilde{s}_j$ is the product of the estimated relative signs along this path. 
If any edge on the path has an inference error, then the estimated sign flips parity for an odd number of errors.  
Assuming (for simplicity) that each edge has the same error probability $p_e \coloneqq  \mathrm{Pr}\left(\tilde{s}_{j,j+2^k}\neq s_{j,j+2^k}\right)$ and errors are independent, the probability $p_j$ that $\tilde{s}_j$ is incorrect can be estimated by a representative value that characterizes the error rate of a root-to-node path
\begin{gather}
    p_j 
    \leq\frac{1-(1-2p_e)^{L}}{2}.
    \label{e:error-propagation-tree}
\end{gather}
In the regime $p_e L \ll 1$, one recovers $p_j\approx p_eL$.
But a moderate $p_e=0.01$ can lead to unreliable predictions deeper in the tree at 10~qubits as $p_j\approx 0.1$.
The effective error rate becomes exponentially smaller by combining $N_{\mathrm{tree}}$ random trees. 
Assuming that each tree has an independent error rate $p_j<1/2$, the probability of returning the wrong signs after majority voting is
\begin{gather}
    \operatorname{Pr}_{\mathrm{HRF}}\left(\tilde{s}_j\neq s_j\right) \leq 
        \exp\left(
            -2N_{\mathrm{tree}}({1}/{2}-p_j)^2
        \right).
    \label{e:majority-voting-prob}
\end{gather}
If each independent tree has an error rate $p_j=0.1$, then $N_{\mathrm{tree}}=11$ trees suffice for a sign error below $3\%$, as shown in~\cref{f:error-bound}~(b). 
If one allows the error budget to be a small fraction of the nodes $\delta$ have sign errors, we require $N_\mathrm{samp.}\geq\ln(L/\delta)/(2\Delta^2)$ to maintain $p_j< 1/2$ and $N_{\mathrm{tree}}\geq \ln(1/\delta)/(2(1/2-p_j)^2)$, where $\Delta =\min_{j,k} \psi_j\psi_{j+2^k}$ is the minimal overlap.

Then, we further analyze the situation of pairwise correlation error between tree $T_i$ and $T_j$ sharing $m$ edges out of a total $2^{N_\mathrm{q}}-1$ edges.
Define the correlation as the ratio of overlapping edges $\gamma= m/\left(2^{N_\mathrm{q}}-1\right)$ and hence covariance can be calculated as $\gamma\cdot p_j(1-p_j)$. 
To interpret the impact of correlation, we introduce the effective number of independent trees
\begin{gather}
    N_{\mathrm{tree}}^{\mathrm{eff}}=\frac{N_{\mathrm{tree}}}{1+\gamma\left(N_{\mathrm{tree}}-1\right)},
    \label{e:corr-tree}
\end{gather}
such that an ensemble of $N_{\mathrm{tree}}^{\mathrm{eff}}$ independent trees would exhibit the same variance as the correlated ensemble of $N_{\mathrm{tree}}$ trees.
Notice in the infinite tree limit, the effective ensemble size approaches $ N_{\mathrm{tree}}^{\mathrm{eff}} \rightarrow 1/\gamma$, regardless of the actual number of trees employed. 
This concentration arises because correlated trees might share common factors that can lead to collective failure.
If $\gamma=1$, all trees become identical copies, and majority voting reduces to a single tree’s prediction with error probability $p_j$.
For large $N_{\mathrm{tree}}$, one can apply the Central Limit Theorem and calculate the majority voting error converging to an error floor 
\begin{gather}
    \mathrm{Pr}_{\infty}=\Phi\left(\frac{p_j-1/2}{\sqrt{\gamma \, p_j\left(1-p_j\right)}}\right),
    \label{e:ctl-error}    
\end{gather}
where $\Phi(\cdot)$ denotes the standard normal cumulative distribution function.
For $p_j=0.1$ and $\gamma=0.15$, one can calculate the bound as $\mathrm{Pr}_{\infty} = 0.03\%$.
The empirical value of $\gamma=0.15$ is obtained by sampling 500 trees at 14 qubits. 
As the problem size increases, $\gamma$ will become even smaller.
This floor represents a fundamental limitation that cannot be eliminated by adding more trees when correlations are present.
In practice, one can assign different sample sizes to the probabilities $\{|\psi_j^k|^2\}$, or introduce randomness into the tree-generation procedure to decorrelate them.

\section{Results}\label{s:result}

\subsection{Simulation}\label{ss:sim-noise}

\begin{figure}[htb]
    \centering
    \includegraphics[]{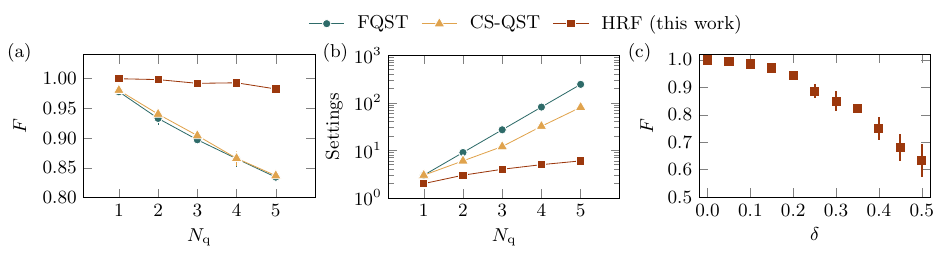}
    \caption{
    Performance and robustness analysis on noisy simulators.
    The state fidelity obtained with FQST, CS--QST, HRF on (a) the noisy simulator mimicking the IBM quantum processor \texttt{ibm\_fez}.
    (b) The number of measurement settings for each method.
    (c) Robustness test of HRF by imaginary perturbations control by $\delta$.
    The data is averaged over 10 different random states with standard deviations plotted as error bars.
    }
    \label{f:simulation-results}
\end{figure}

We first benchmark the performance of FQST, CS--QST, and HRF on the Qiskit simulator~\citep{qiskit2024}, targeting random real-valued pure states, with $N_\mathrm{samp.}=10^4$ samples per circuit evaluation.
For CS--QST, we use an empirical measurement setting scaling of $N_\mathrm{q} \, 2^{N_\mathrm{q}}/2$ and achieve similar performance with FQST under noisy simulation.
The tunable weight $\lambda$ defined in the trace-norm optimization \cref{e:qst-cs} is chosen as $\lambda=0.1$.
For HRF, we use $N_{\mathrm{tree}}=111$ to ensure high-fidelity reconstruction.
The results are reported in \cref{f:simulation-results}.
The proposed method consistently shows the highest fidelity with linear scaling of measurement settings.
For the detailed hardware noise modeling in \cref{f:simulation-results} (a), see Section~4.2~\citep{song2025incompressible}.
Here, we emphasize that typical incoherent quantum noise (such as depolarizing and phase-flip noise) does not violate the real-value assumption. 
Coherent error, such as small-angle over-rotations $R_z$, can, in principle, introduce imaginary components.
However, the probability of such an error is much lower than that of other dominant error sources, since $R_z$ rotation is physically implemented via a virtual gate on IBM hardware~\citep{mckay2017efficient}.

We further test the robustness of HRF by injecting small imaginary components into the state vector as $|\psi(\delta)\rangle = \left(|\psi\rangle + i\delta |\phi\rangle \right)/\mathcal{N}_\psi$, where $|\psi\rangle$ are the real-valued pure states to reconstruct, $\mathcal{N}_\psi$ is the normalization constant, $|\phi\rangle$ is the perturbation following isotropic multivariate Gaussian distribution, and $\delta$ controls the perturbation strength.
The reconstruction fidelity $F = |\langle\widetilde{\psi(\delta)}|\psi\rangle|^2$ where $|\widetilde{\psi(\delta)}\rangle$ stands for the HRF reconstruction of the perturbed state, is shown in \cref{f:simulation-results} (c) with $N_\mathrm{q}=8$ and $N_{\mathrm{tree}}=111$.
HRF demonstrates strong noise resilience under small perturbation $\delta\leq 0.2$ while maintaining fidelity $F\geq 90\%$.

\subsection{Benchmarks on IBM quantum hardware}\label{ss:ibm-experiment}

\begin{figure}[htb]
    \centering
    \includegraphics[]{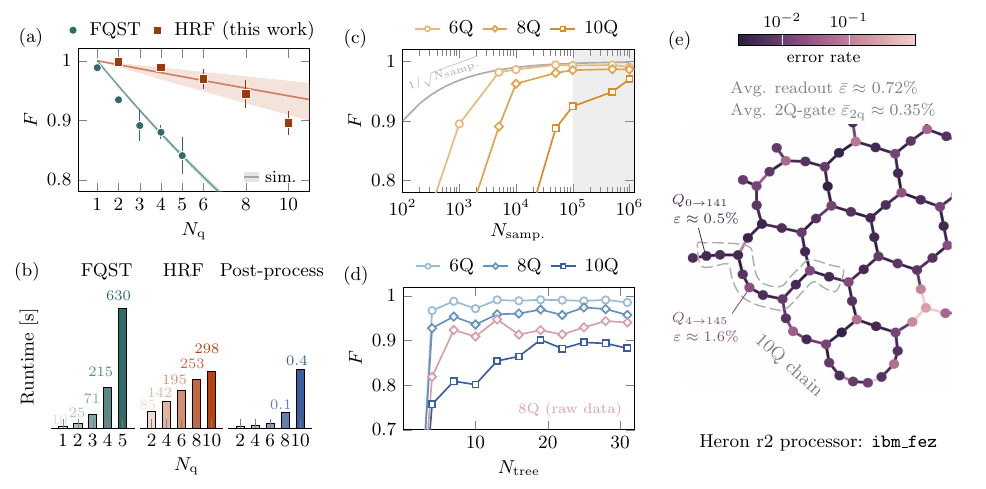}
    \caption{
        Performance and runtime analysis on IBM quantum hardware.
        (a) The fidelities of the reconstructed state from FQST and HRF method running on \texttt{ibm\_fez} quantum processor. 
        Error bars indicate the standard deviation of the hardware results, and the shaded area represents the variance.
        (b) Runtime (seconds) scaling of FQST, HRF on \texttt{ibm\_fez} and post-processing time using $111$ trees. 
        (c) The impact of samples $N_\mathrm{samp.}$ to fidelities on noisy simulator. 
        The solid gray line represents the upper bound limited by finite sampling noise $\epsilon \approx 1/\sqrt{N_\mathrm{samp.}}$ and $N_\mathrm{samp.} = \num{1e5}$ is the maximal sample size used in experiments. 
        (d) The impact of the number of trees $N_\mathrm{tree}$ used in post-processing for sign determination based on samples with and without error mitigation. 
        (e) Readout error $\varepsilon$ and 2~qubit gate error $\varepsilon_\mathrm{2q}$.
    }
    \label{f:hardware-result}
\end{figure}

We demonstrate the HRF method and compare its performance and runtime scaling with standard FQST in the Pauli basis on \texttt{ibm\_fez}, the IBM~Heron~r2 superconducting quantum processor.
\Cref{f:hardware-result}~(e) shows the readout and native two-qubit gate (CZ) error rates.
We select a 10-qubit chain on the device to reduce readout error, which is the dominant error source.  
The real-valued quantum states were generated using the hardware-efficient ansatz~\citep{kandala2017hardware}, but the gate set is limited to $R_y(\theta_j)$ rotation gates on each qubit and $\mathrm{CNOT}_{j,j+1}$ on adjacent qubits.
To minimize the influence of two-qubit gate errors, we use a constant shallow depth state preparation circuit $U_{\mathrm{prep.}}$ that consists of $4$ layers of $R_y$ gates and $4(N_\mathrm{q}-1)$ pairwise $\mathrm{CNOT}$ gates inside. 
We use measurement-error mitigation for the FQST and HRF experiments by inverting an assignment matrix~\citep{nation2021scalable}.
We adopt the $X_{+}X_{-}$ dynamical decoupling sequence to suppress decoherence during qubit idling time~\citep{tripathi2022suppression, niu2022effects, ezzell2023dynamical}. 

For FQST cases, we reconstruct the density matrix $\rho$ up to 5~qubits by measuring $3^{N_\mathrm{q}}$ global observables with $N_\mathrm{samp.}=10^4$ samples for each observable.
The 5~qubit FQST experiment takes about $\SI{10}{\minute}$ to measure $3^5=243$ observables on \texttt{ibm\_fez}.
In contrast, HRF requires only $\SI{18}{\second}$ of sampling time for the same problem size.
The experiments were performed sequentially, with periodic calibrations performed without interruption. 
Thus, the total running time, including error mitigation overhead, accurately reflects the circuit execution time, providing a faithful measure of the protocol's efficiency and the experimental resources required. 

Performing FQST for up to 10~qubits can be time-consuming ($\sim \SI{6}{\hour}$) on quantum hardware. 
For HRF, we reconstruct the state vector $|\psi\rangle$ up to 10~qubits by sampling $N_\mathrm{q}+1=11$ circuits with $N_\mathrm{samp.}=10^5$ samples for $N_\mathrm{q}>5$ and $N_\mathrm{samp.}=10^4$ samples for $N_\mathrm{q}\leq 5$ to ensure fair fidelity comparison with FQST.
The transpiled 10~qubit circuit is $\SI{2.54}{\micro\second}$ long and the average $T_2$ coherence time on the 10~qubit chain of \texttt{ibm\_fez} is $\SI{135.74}{\micro\second}$. 
We provide more detailed hardware calibration in Supplementary Note \ref{sn:hrf}.
Noisy simulations conducted in this and later sections use the same calibration data.

First, we confirm the feasibility of HRF for real-valued pure-state reconstruction and demonstrate its superiority over FQST in terms of accuracy and runtime scaling.
\Cref{f:hardware-result}~(a) shows the fidelity $F(|\psi\rangle,|\tilde{\psi}\rangle) = |\langle\psi|\tilde{\psi}\rangle|^2$ between the target state $|\psi\rangle$ and reconstructed state from hardware $|\tilde{\psi}\rangle$.
The average fidelity among $R=5$ random states reaches $84.05\%$ for 5~qubit FQST and $89.53\%$ for 10~qubit HRF after \num{2.43e6} and \num{1.10e6} total samples, respectively.
These experimental results are consistent with the noisy simulations shown in the shaded region using $R=10$ random states. 
HRF exhibits higher average fidelity with fewer samples, but larger variance due to post-processing randomness. 
So, the reconstructed state has a smaller statistical bias with respect to the true fidelity of $1$, but may exhibit higher variance due to fewer measurements (shots).
This bias--variance trade-off is common in quantum state tomography on noisy devices~\citep{yang2023experimental}. 
In practice, the community often prioritizes reducing bias over variance, as bias represents a systematic error, whereas variance can be suppressed by averaging over more samples~\citep{schwemmer2015systematic}.

\Cref{f:hardware-result}~(b) presents the linear runtime scaling for sample collection using $N_\mathrm{samp.}=10^5$ and a $\Omega(2^{N_\mathrm{q}})$ scaling in post-processing, where the latter part can leverage parallel computing to distribute the workload on multiple CPUs since $\mathcal{Q}_{N_\mathrm{q}}$ is always bipartite~\citep{harary1988survey}.
The largest FQST reconstruction takes $\SI{3.35}{\hour}$ post-processing time for a 14-qubit state~\citep{hou2016full} while HRF would take only $\SI{11}{\second}$ post-processing time for the same problem size using $N_{\mathrm{tree}}=300$.

We show the impact of sample size $N_\mathrm{samp.}$ and the number of trees $N_{\mathrm{tree}}$ in \cref{f:hardware-result}~(c) and~(d) via simulations.
The $10$~qubit HRF fidelity could reach $97.05\%$ using $N_\mathrm{samp.}=10^6$ samples on a noisy simulator.
As the state size increases, more samples are required to reach the finite sampling limit, as indicated by the solid gray curve. 
We observe that as more samples are collected, the estimated fidelity converges rapidly and asymptotically approaches the limit set by the sampling noise. 
This suggests that fluctuations in protocol performance are primarily due to statistical noise, with no additional sources of error, indicating the protocol's efficiency and stability.
For post-processing, the reconstructed state fidelity converges after doing majority voting using $N_\mathrm{tree}\approx30$ spanning trees on the corresponding hypercube graph.
As illustrated in the pink curve (raw samples without error mitigation), higher-fidelity sampling with measurement error mitigation would reduce the final reconstructed state fidelity by $\sim3\%$.
The overhead introduced is a few extra hardware calibrations, which require approximately \SI{10}{\second} for 10~qubit states.

\subsection{Quantum state property and overlap estimation}\label{ss:property-estimation}

\begin{figure}[htb]
    \centering
    \includegraphics[]{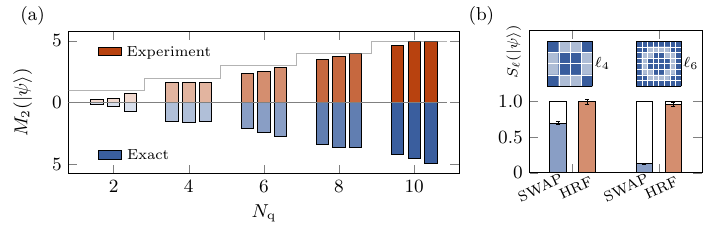}
    \caption{
        Estimating quantum state properties using the reconstructed state from HRF.
        (a) Stabilizer entropy of $R=3$ random states on \texttt{ibm\_fez}.
        (b) State overlap (noisy simulation) that resembles a numerical integration over a circle in the domain, where $\ell_4$ and $\ell_6$ are integration paths for 4- and 6-qubit.
    }
    \label{f:state-property}
\end{figure}

In this section, we apply the HRF method to evaluate nonlinear state properties, including $\alpha$-stabilizer R\'enyi entropy $M_\alpha(|\psi\rangle)$ for quantifying quantum magic~\citep{leone2022stabilizer}, and overlap $S_\ell(|\psi\rangle)$ between a classical index state $|\psi_\ell\rangle$ that resembles a numerical integration along the path $\ell$.
The common magic measure $\alpha$-stabilizer R\'enyi entropy~\citep{leone2022stabilizer} is defined as
\begin{gather}
    M_\alpha(|\psi\rangle)=\frac{1}{1-\alpha} \log_2 \left(\sum_{\sigma \in \mathcal{P}} 2^{-N_\mathrm{q}}\langle\psi| \sigma|\psi\rangle^{2 \alpha}\right),
    \label{e:quantum-magic}
\end{gather}
where $\alpha$ is the entropic index and $\mathcal{P}$ is the set of $4^{N_\mathrm{q}}$ Pauli strings.
Magic states are the resource that allows quantum computers to attain an advantage over classical computers. 
Given its importance, we measure the 2-stabilizer entropy $M_2(|\tilde{\psi}\rangle)$ of $3$ random real-valued states on \texttt{ibm\_fez}, as shown in \cref{f:state-property}~(a).
The results closely match an exact evaluation.

We use HRF to estimate the overlap $S_\ell(|\psi\rangle) = |\langle\psi_{\ell_{N_\mathrm{q}}}|\psi\rangle|^2$ where $\{\ell_{N_\mathrm{q}}\}$ is a circle path in a 2D lattice of size $[2^{N_\mathrm{q}/2}, 2^{N_\mathrm{q}/2}]$  as illustrated in \cref{f:state-property}~(b), and 
$|\psi_{\ell_{N_\mathrm{q}}}\rangle=\frac{1}{\mathcal{N}}\sum_{j \in \{\ell_{N_\mathrm{q}}\}}|j\rangle$
is the corresponding state by concatenating the lattice row-wise for a normalization constant $\mathcal{N}$.
This quantity resembles a numerical integration over the path $\ell_{N_\mathrm{q}}$, which can occur, for example, when calculating drag and lift coefficients for computational fluid mechanics simulations~\cite{penuel2024feasibility}, scattering cross sections in electromagnetism simulations~\cite{clader2013preconditioned}, among others.
Through noisy simulations, we compare the estimated overlap with the standard SWAP test~\citep{buhrman2001quantum}.
The SWAP test requires $2N_\mathrm{q}+1$~qubits in total, and the explicit circuit construction with estimated hardware execution time is shown in Supplementary Note \ref{sn:measure}.
The accuracy of the SWAP test decreases with increasing problem size $N_\mathrm{q}$ due to hardware noise, whereas HRF maintains a high fidelity of $\geq 95\%$ (averaged over $R=5$ random states). 

\subsection{Quantum linear system solver}\label{ss:qlsa}

\begin{figure}[htb]
    \centering
    \includegraphics[]{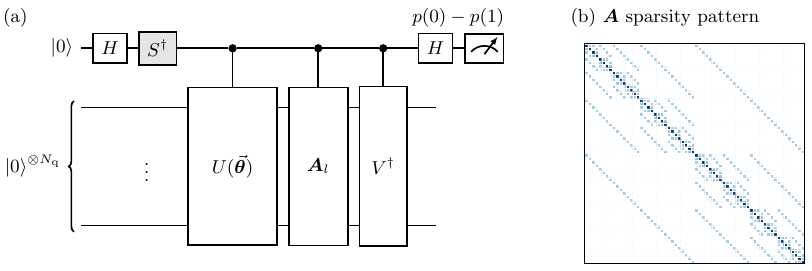}
    \caption{
        Visualization of the quantum linear solver circuit and problem matrix. 
        (a) VQLS circuit for evaluating the cost function $C(\btheta)$ with the Hadamard test, which uses one ancilla qubit initialized in $|0\rangle$ and a controlled unitary sandwiched with two Hadamard gates.
        Then, one can calculate $\operatorname{Re}\langle 0| V^{\dagger} \bA_l U(\btheta)|0\rangle = p(0) - p(1)$, where $p(0)$ and $p(1)$ are the probabilities of measuring $0$ and $1$ in the ancilla qubit.
        (b) Visualization of the matrix structure of $\bA_{64\times 64}$ in a 6-qubit Ising-inspired linear system problem.
    }
    \label{f:vqls-matrix}
\end{figure}

In this section, we first use the variational quantum linear solver (VQLS)~\citep{bravo2023variational} shown in \cref{f:vqls-matrix}~(a) to solve a linear system $\bA \ket{x} = \ket{b}$ inspired by the Ising model under noiseless simulation. 
Then, we apply the HRF method to read out the solution state and benchmark its performance on Heron r2 \texttt{ibm\_fez} and Heron r3 \texttt{ibm\_boston}. 
The Ising-inspired quantum linear system problem~\citep{bravo2023variational}  can be written as
\begin{gather}
    \bA = \frac{1}{\zeta} \left(\sum_{k=0}^{N_\mathrm{q}-1} \sigma_x^{k} + J\sum_{k=0}^{N_\mathrm{q}-2} \sigma_z^{k}\sigma_z^{k+1} + \eta\mathbb{I}\right)
    \quad \text{and} \quad
    \ket{b} = \bigotimes_{k=0}^{N_\mathrm{q}-1} R_y(\theta^{k})\ket{0}^{\otimes N_\mathrm{q}},
    \label{e:ising-qlsa}
\end{gather}
where $J$ is the coupling strength, $\zeta$ and $\eta$ are normalization factors to control the matrix condition number $\kappa(\bA):=\|\bA\|\left\|\bA^{-1}\right\|$ such that the smallest eigenvalue of $\bA$ is $1/\kappa$ and largest eigenvalue of $1$.
Under this evaluation, we set $J=0.1$ and $\kappa\in\{2, 10\}$.
The sparsity pattern of matrix $\bA$ is illustrated in \cref{f:vqls-matrix}~(b).
The state preparation circuit $V$ for the right-hand side vector $\ket{b}$ consists of one layer of $R_y$ rotation on each qubit, and the rotation angles $\theta^{k}$ are chosen randomly from $[-\pi,\pi]$ applied on the $k$-th qubit.

The VQLS algorithm essentially uses the Hadamard test~\citep{aharonov2006polynomial} to estimate the cost function $C(\btheta) = 1 - |\langle b|\Psi\rangle|^2$, where $|\Psi\rangle = \bA|\psi\rangle/\sqrt{\langle \psi| \bA^{\dagger} \bA| \psi \rangle}$ and $|\psi\rangle = U(\btheta)|0\rangle^{\otimes N_\mathrm{q}} $ is the quantum iterative solution prepared using the same hardware-efficient ansatz as in \cref{ss:ibm-experiment}.
Hence, the cost function has an operational meaning similar to the absolute residual $\||b\rangle-\bA|\psi\rangle\|$.
Then, one can optimize over the parameters $\btheta = (\theta_1, \dots, \theta_m)$ to minimize the cost function and obtain an approximate solution $\ket{\psi^*}$.
It is usually assumed that matrix $\boldsymbol{A}$ is decomposed into a linear combination of unitaries (LCU) as
\begin{equation}
    \bA =\sum_{l=1}^{L_a} c_l \bA_l.
\end{equation}
For the Ising-inspired linear system problem in \cref{e:ising-qlsa}, each $\bA_l$ is a Pauli operator and $c_l\in\mathbb{R}$.

\begin{figure}[htb]
    \centering
    \includegraphics[]{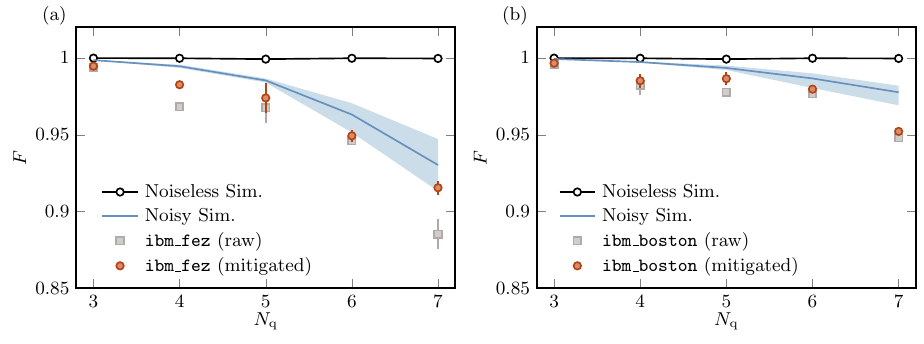}
    \caption{
        Quantum linear system solves on IBM quantum hardware.
        Reconstructing the solution state from a variational quantum linear solver on (a) \texttt{ibm\_fez} and (b) \texttt{ibm\_boston} (average 2Q-gate error $\bar{\varepsilon}_\mathrm{2q}\approx 0.11\%$ and readout error $\bar{\varepsilon} \approx 0.51\%$).
        The raw fidelities are shown in gray squares and error mitigated results are shown in rust dots.
        Error bars indicate the standard deviation of the hardware results, and the blue shaded area represents the variance of noisy simulations.
    }
    \label{f:vqls-hrf}
\end{figure}

After solving the linear system, we apply the HRF method to read out the solution state $\ket{\psi^*}$ for $N_\mathrm{q} \in \{3,4,5,6,7\}$ qubits with the noiseless simulation, noisy simulation based on hardware noise modeling, and actual quantum hardware \texttt{ibm\_fez} with $\kappa=2$ and \texttt{ibm\_boston} with $\kappa=10$.
The results measuring the fidelity between the VQLS solution $\ket{\psi^*}$ and the reconstructed state by the HRF method $\ket{\tilde{\psi}}$ are shown in \cref{f:vqls-hrf}.
The hardware results are consistent with noisy simulation behavior, and one can observe the performance improvement from error mitigation.
HRF reconstructs the 7-qubit solution state with high fidelity $F=91.55\%$ on \texttt{ibm\_fez} and $F=95.21\%$ on \texttt{ibm\_boston} by using \num{8e5} total samples.

\section{Conclusion}\label{s:conclusion}

We present an asymptotically faster technique for reconstructing real-valued pure quantum states.
HRF reduces the number of measurement settings for pure-state reconstruction from exponential to linear in qubit count.
Still, the post-processing time remains exponential, though it can be mitigated via parallel computing on classical devices.
We experimentally demonstrate our approach on an IBM superconducting quantum processor.
Using the superposition of amplitudes and a random forest algorithm as a post-processing tool, we reconstruct the random 10~qubit real-valued states with high fidelity $F\approx89\%$ and a 7~qubit solution state from the Ising-inspired quantum linear system problem with fidelity $F\approx 95\%$.
HRF accurately estimates state properties, such as quantum magic and state overlap, from near-term hardware sampling results.
This improvement paves the way for end-to-end applications based on QLSA and for many application problems on near-term, early fault-tolerant devices.
We want to emphasize that HRF is not only targeting near-term quantum processors but also fault-tolerant devices where the pure real-valued assumption can be reliably held with the protection of quantum error correction codes.
The method may also shed light on other quantum tomography techniques, such as real-valued process tomography and more efficient reconstruction of mixed real-valued states.

\section*{Data and code availability}

The code and data used for this manuscript are permissively (MIT) licensed and available at \url{https://github.com/comp-physics/Quantum-HRF-Tomography}.

\section*{Competing interests}

The authors declare no competing interests.

\section*{Acknowledgments}

We acknowledge the use of IBM Quantum services for this work.
The views expressed are those of the authors and do not reflect the official policy or position of IBM or the IBM~Quantum team.
This research used resources of the Oak Ridge Leadership Computing Facility, which is a DOE Office of Science User Facility supported under Contract DE-AC05-00OR22725. 

\section*{Author Contributions}
N. R. formulated the initial idea; Z. S. and H. R. developed the theory; 
Z. S. and M. L. conducted numerical simulations and performed experiments on IBM hardware;
S. B. and B. G. supervise the project and manuscript writing.
All co-authors contributed to the preparation of the manuscript.

\bibliographystyle{bibsty}
\bibliography{main.bib}

\renewcommand{\theequation}{\arabic{equation}}
\renewcommand{\thefigure}{S\arabic{figure}}
\renewcommand{\thetable}{S\arabic{table}}
\renewcommand{\thesection}{\arabic{section}}
\renewcommand{\thesubsection}{\arabic{section}.\arabic{subsection}}
\renewcommand{\thesubsubsection}{\arabic{section}.\arabic{subsection}.\arabic{subsubsection}}
\setcounter{equation}{0}
\setcounter{table}{0}
\setcounter{section}{0}
\setcounter{figure}{0}

\clearpage
\newpage
\vspace{2cm}
\onecolumngrid
\vspace{2cm}
\begin{center}
    {\textbf{\large Supplemental Material}} 
\end{center}

\section{SUPPLEMENTARY NOTE 1: Details of HRF}\label{sn:hrf}

\subsection{Algorithm description via pseudocode}

\begin{algorithm}[H]
\caption{Hadamard Random Forest (HRF)}\label{alg:hrf}
\begin{algorithmic}[1]
\Require Number of qubits $N_\mathrm{q}$, sample size for each circuit $N_\mathrm{samp.}$, number of trees $N_\mathrm{tree}$
\Ensure  Reconstruct real state vector $|\psi\rangle \in\mathbb{R}^{2^{N_\mathrm{q}}}$

\Statex
\Comment{1. \textbf{Collect samples}}
\State Prepare $N_\mathrm{samp.}$ copies of $|\psi\rangle$ using $U_{\mathrm{prep.}}$
\State Measure all $N_\mathrm{q}$ qubits in $\sigma_z$ basis
\ForAll{$j\in\{0,\dots,2^{N_\mathrm{q}}-1\}$}
  \State $|\tilde{\psi}_j|^2\gets$ empirical frequency in basis $|j\rangle$
\EndFor
\For{$k=0$ to $N_\mathrm{q}-1$}
  \State Prepare $N_\mathrm{samp.}$ copies of $|\psi\rangle$ using $U_{\mathrm{prep.}}$ 
  \State Measure the $(N_\mathrm{q}-k-1)$-th qubit in $\sigma_x$ basis 
  \State Measure remaining qubits in the $\sigma_z$ basis
  \ForAll{$j\in\{0,\dots,2^{N_\mathrm{q}}-1\}$}
    \State $|\tilde{\psi}^k_j|^2\gets$ empirical frequency in basis $|j\rangle$
  \EndFor
\EndFor

\Statex
\Comment{2. \textbf{Build random forest}}
\For{$i=1$ to $N_\mathrm{tree}$}
  \State $T_i \gets$ a random spanning tree on $\mathcal{Q}_{N_\mathrm{q}}$ using BFS
  \State Initialize root sign $s^{(i)}_0 \gets +1$
  \ForAll{unique path from node $0\to j$ in $T_i$}
    \ForAll{valid $(n,k)$ along path}
        \State Compute relative signs 
        \State $s^{(i)}_{n,n+2^k} \gets \sgn(2|\tilde{\psi}^k_n|^2-|\tilde{\psi}_n|^2-|\tilde{\psi}_{n+2^k}|^2)$
        \State $s^{(i)}_j \gets s^{(i)}_{j,j-2^k}\times \cdots \times s^{(i)}_{2^k,0} \times s^{(i)}_0$ 
        \EndFor
  \EndFor
\EndFor

\Statex
\Comment{3. \textbf{Majority voting}}
\ForAll{$j\in\{0,\dots,2^{N_\mathrm{q}}-1\}$}
  \State $s_j \gets \sgn\Bigl(\sum_{i=1}^{N_\mathrm{tree}} s^{(i)}_j\Bigr)$
\EndFor

\Statex
\Comment{4. \textbf{Reconstruct state}}
\ForAll{$j\in\{0,\dots,2^{N_\mathrm{q}}-1\}$}
  \State $|\tilde{\psi}_j| \gets \sqrt{|\tilde{\psi}_j|^2}$
  \State $\tilde{\psi}_j \gets s_j\times|\tilde{\psi}_j|$
\EndFor
\State \Return $|\tilde{\psi}\rangle$ 
\end{algorithmic}
\end{algorithm}

\subsection{Quantum circuit construction}

\begin{figure}[htb]
    \centering
    \includegraphics[scale=1]{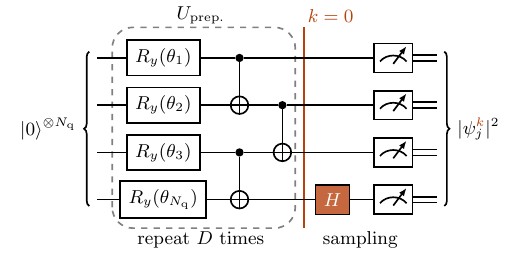}
    \caption{State preparation and measurement circuits for HRF.
    The largest case on the IBM hardware benchmark \cref{ss:ibm-experiment} contains 36~CZ gates.
    }
    \label{f:spam-cir}
\end{figure}

We use the hardware-efficient ansatz to prepare real-valued states as shown in \cref{f:spam-cir}. 
The parameters are chosen uniformly from the interval $[-\pi/2, \pi/2]$.

\subsection{Hardware calibration data}\label{app:calibration}

\Cref{tab:hardware-metric} shows the hardware calibration data used throughout the manuscript.

\begin{table}[htb]
    \centering
    \begin{tabular}{@{} c  c  c  c  c @{}}
        \toprule
        Qubit & $T_1$ (\unit{\micro\second}) & $T_2$ (\unit{\micro\second})   & Meas.\ error $\varepsilon$ & 2Q gate error $\varepsilon_{\mathrm{2q}}$ \\
        \midrule
        $Q_{0}$  &  184  &  101  &  \num{4.64e-3}  &  \num{2.74e-3}  \\
        $Q_{1}$  &  188  &  237  &  \num{4.88e-3}  &  \num{2.69e-3}  \\
        $Q_{2}$  &  216  &  173  &  \num{4.39e-3}  &  \num{3.02e-3}  \\
        $Q_{3}$  &  171  &  248  &  \num{6.34e-3}  &  \num{3.35e-3}  \\
        $Q_{4}$  &  142  &  111  &  \num{1.56e-2}  &  \num{3.35e-3}  \\
        $Q_{5}$  &  197  &  \phantom{1}31  &   \num{5.62e-3}  &  \num{3.48e-3}  \\
        $Q_{6}$  &  170  &  152  &  \num{9.28e-3}  &  \num{5.54e-3} \\
        $Q_{7}$  &  140  &  138  &  \num{4.15e-3}  &  \num{3.42e-3}  \\
        $Q_{8}$  &  154  &  111  &  \num{1.29e-2}  &  \num{3.65e-3}  \\
        $Q_{9}$  &  \phantom{1}98   &  \phantom{1}57   &  \num{4.64e-3}  &  /  \\
        \midrule
        Avg. &  166  &  136  &  \num{7.24e-3}  &  \num{3.47e-3}  \\
        \bottomrule
        \end{tabular}
    \caption{
        Device properties of \texttt{ibm\_fez} at the time each experiment reported in this paper was performed.
        The readout length is $t_{\mathrm{mea.}} = \SI{1560}{\nano\second}$ and native 2Q gate (CZ) with pulse length $t_{\mathrm{2q}}=\SI{84}{\nano\second}$ for all qubits.
    }
    \label{tab:hardware-metric}
\end{table}

\section{SUPPLEMENTARY NOTE 2: MEASUREMENT SCHEMES}
\label{sn:measure}

\subsection{Quantum Amplitude Estimation}

Compared to quantum amplitude estimation (QAE)~\citep{grinko2021iterative,manzano2023real}, HRF employs shallow circuits with easily implementable gates.
QAE is optimal for estimating a specific amplitude or a few dominant amplitudes.
Still, it requires deep, coherent circuits with controlled unitaries, which are challenging to implement and prone to errors.
HRF offers a more scalable alternative that recovers all amplitudes simultaneously, rather than running QAE sequentially, making it well-suited for near-term quantum applications.
\cref{tab:qae-vs-hrf} provides a detailed comparison of HRF and QAE.

\begin{table}[htb]
  \centering
  \setlength{\tabcolsep}{6pt} 
  \begin{tabular}{@{} l ll @{}}
    \toprule
    Feature          & QAE  & HRF (this work) \\
    \midrule
    Objective
      & \makecell[l]{Estimate single amplitude of general\\ complex quantum states}
      & \makecell[l]{Reconstruct full state vector of\\ real-valued quantum states} \\

    Circuit Depth
      & Deep (uses Grover-like subroutines)
      & Shallow (single-layer Hadamards) \\

    Measurements
      & Controlled unitaries with ancilla
      & $\sigma_z,\sigma_x$ only \\

    Qubit Overhead
      & $N_\mathrm{q}+1$ or more (ancilla + control)
      & $N_\mathrm{q}$ \\

    Sample Complexity
      & Optimal $\mathcal{O}(1/\epsilon)$
      & Trade more samples for shallow circuits \\

    NISQ Suitability
      & Low; Error-prone
      & High; Error-robust \\
    \bottomrule
  \end{tabular}
  \caption{Comparison between Quantum Amplitude Estimation and Hadamard Random Forest.}
  \label{tab:qae-vs-hrf}
\end{table}

\subsection{Quantum state overlap estimation}

\begin{figure}[htb]
    \centering
    \includegraphics[]{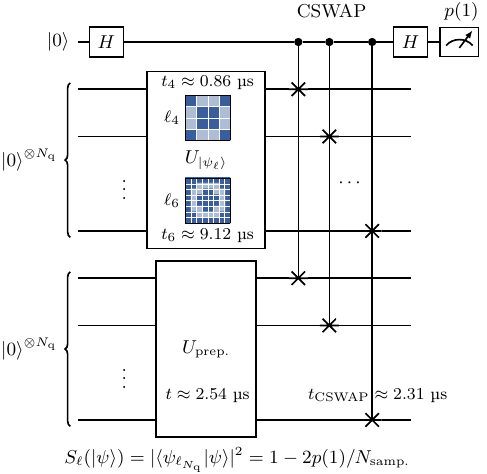}
    \caption{SWAP test for estimating overlap $S_\ell(\ket{\psi})$ with transpiled execution time on \texttt{ibm\_fez} for each block.
    }
    \label{f:swap-test}
\end{figure}

Here, we show a concrete example of how HRF can simplify the task of estimating the overlap $S = |\langle \psi|\phi \rangle|^2$ between two unknown quantum states $|\psi\rangle$ and $|\phi\rangle$.
Such a task has applications in quantum machine learning~\cite{havlivcek2019supervised, liu2021rigorous} and cross-platform verification~\cite{elben2020cross}.
A widely adopted approach to this problem is the SWAP test~\cite{buhrman2001quantum}, which involves applying a controlled-SWAP operation and measuring an ancillary qubit (see \cref{f:swap-test}).
However, the SWAP test requires direct preparation and entanglement between the two states, making its implementation resource-intensive on near-term devices.

The HRF offers an alternative method for estimating state overlap.
Instead of requiring entanglement between $|\psi\rangle$ and $|\phi\rangle$, HRF reconstructs each state independently and computes their overlap classically via post-processing of reconstructed amplitudes. 
This approach significantly reduces the circuit size by half.
It also eliminates the need for controlled-SWAP operations, making it more scalable and easier to implement on near-term quantum hardware.
In larger circuits, like the SWAP test, noise accumulates.
Since HRF uses smaller circuits, it is more resilient to noise and can integrate well-established error-mitigation methods, such as SPAM error cancellation, as demonstrated in this work.
The lower circuit complexity also enables the use of other noise-reduction techniques.

Regarding sample complexity and scalability, suppose there are $R$ states.
The swap test requires circuits involving all $R^2$ pairs of states, whereas HRF only needs to characterize each of the $R$ states individually, yielding a quadratic speedup in the number of states.
We note, however, that HRF shifts the expensive cross-state quantum operations into classical post-processing, at the cost of potentially more measurement shots.
Thus, HRF is more advantageous for near-term devices where minimizing quantum circuit depth is crucial.
Beyond state overlap estimation, the advantage brought by HRF can also generalize to other tasks involving multi-state property estimation, such as entanglement estimation~\cite{mintert2005concurrence, ekert2002direct}.

In this study, we apply the SWAP test to estimate overlap that resembles a numerical integration over the path $\ell_{N_\mathrm{q}}$.
More specifically, the index state that encodes the 4-qubit path $\ell_4$ is shown below
\begin{align}
    |\psi_{\ell_4}\rangle = &\frac{1}{2\sqrt{2}} \big{(}\ket{0001}+\ket{0010}+\ket{0100}+\ket{0111} \notag \\
        +&\ket{1000}+\ket{1011}+\ket{1101}+\ket{1110} \big{)}.
    \label{e:index-state-4q}
\end{align}
This happens to be a stabilizer state with $M_2(|\psi_{\ell_4}\rangle)=0$ but more generally $|\psi_{\ell_6}\rangle$ is not.

\end{document}